\documentclass[runningheads]{llncs}
\usepackage{graphicx}
\usepackage{amsmath}
\usepackage{todonotes}
\usepackage{multirow}
\usepackage{graphicx}
\usepackage{epstopdf}
\AppendGraphicsExtensions{.tif}

\begin{document}
\title{OUTCOMES: Rapid Under-sampling Optimization achieves up to $50\%$ improvements in reconstruction accuracy for multi-contrast MRI sequences}
%
%
\author{Ke Wang \inst{1}\and
Enhao Gong\inst{2} \and
Yuxin Zhang\inst{3,4} \and Suchadrima Banerjee\inst{5} \and Greg Zaharchuk\inst{6} \and John Pauly\inst{7}}
\institute{Electrical Engineering and Computer Sciences, UC Berkeley, Berkeley, CA, USA \email{kewang@berkeley.edu} \and Subtle Medical Inc., Menlo Park, CA, USA \and Medical Physics, University of Wisconsin-Madison, Madison, Wisconsin, USA \and Radiology, University of Wisconsin-Madison, Madison, WI, US \and Electrical Engineering, Stanford University, Stanford, CA, USA \and Radiology, Stanford University, Stanford, CA, USA }
%
\maketitle              
\begin{abstract}
Multi-contrast Magnetic Resonance Imaging (MRI) acquisitions from a single scan have tremendous potential to streamline exams and reduce imaging time. However, maintaining clinically feasible scan time necessitates significant undersampling, pushing the limits on compressed sensing and other low-dimensional techniques. During MRI scanning, one of the possible solutions is by using undersampling designs which can effectively improve the acquisition and achieve higher reconstruction accuracy. However, existing undersampling optimization methods are time-consuming and the limited performance prevents their clinical applications. In this paper, we proposed an improved undersampling trajectory optimization scheme to generate an optimized trajectory within seconds and apply it to subsequent multi-contrast MRI datasets on a per-subject basis, where we named it OUTCOMES. By using a data-driven method combined with improved algorithm design, GPU acceleration, and more efficient computation, the proposed method can optimize a trajectory within 5-10 seconds and achieve $30\%$ - $50\%$ reconstruction improvement with the same acquisition cost, which makes real-time under-sampling optimization possible for clinical applications.

\keywords{Magnetic Resonance Imaging (MRI)  \and Trajectory optimization \and Parallel imaging \and Compressed sensing \and Multi-contrast MRI sequence.}
\end{abstract}
%
%
%
\section{Introduction}
Multi-contrast Magnetic Resonance Imaging (MRI), such as T1-weighted (T1w), T2-weighted (T2w) and Diffusion-weighted imaging (DWI), has been widely applied to clinical routines because of their different tissue contrast and high lesion sensitivity comparing with single contrast imaging. Traditional clinical brain protocols separately acquire images from different contrasts, which will inevitable suffer from long scanning time. In recent years, there is an increased research interest in acquiring or synthesizing more multi-contrast images to facilitate diagnosis from a single scan, which have shown great improvement in this problem. \cite{MAGiC,barentsz2012esur,kitajima2010prostate,langer2009prostate,tamir2017t2,tamir2020computational,wang2020high}. However, for most of these methods, long scanning time and reconstruction time still limits their clinical applications.

In this regard, for more widely clinical applications of multi-contrast MRI, shorter acquisition time is highly desired. Currently, each sequence among multi-contrast acquisitions has its own acceleration strategy separately with either parallel imaging (PI)\cite{lustig2010spirit,uecker2014espirit} or compressed sensing (CS) \cite{lustig2007sparse}. Besides, different patients will undergo the same imaging protocol and undersampling patters regardless of their different imaging quality. Under this current separate acceleration strategy, limited reduction factors (R) can be achieved to ensure the image quality of different contrasts for different patients. In order to achieve higher reduction factors while maintaining diagnostic image quality, a variety of methods have been proposed to optimize the undersampling trajectory \cite{huang2014fast,bilgic2011multi,levine2018fly}. Although these algorithms are able to improve the resulting reconstruction accuracy compared to a fixed uniform undersampling trajectory at the same reduction factor, the high computational costs and limited performance of existing methods prevent their clinical applications. At the same time, most of these algorithms also requires repeated PI+CS image reconstruction using different undersampling proposals, which is time-consuming by itself. 

Therefore, in this study, we proposed a novel and efficient undersampling trajectory optimization scheme to generate an optimized trajectory within seconds and apply for subsequent multi-contrast MRI datasets on a per-subject basis, mainly for Cartesian sampling but can be easily extendable for non-cartesian cases. We named it as OUTCOMES. Specifically, we proposed a new objective function to approximate the performance of any undersampling pattern without explicitly iterative PI + CS reconstruction, which makes the scheme more efficient and results in better reconstruction performance. By using a data-driven method combined with improved algorithm design, Graphics Processing Unit (GPU) acceleration and more efficient computation, the proposed method can optimize a trajectory within 5-10 seconds and achieve $30\%$ to $50\%$ reconstruction improvement with the same acquisition cost, which finally makes real-time under-sampling optimization possible for clinical applications. Finally, in order to test our algorithm, OUTCOMES is applied to brain imaging and prostate imaging of the human to demonstrate the effectiveness of our proposed method and feasibility to clinical applications.

\begin{figure}[ht!]
\centering
\includegraphics[width=10cm]{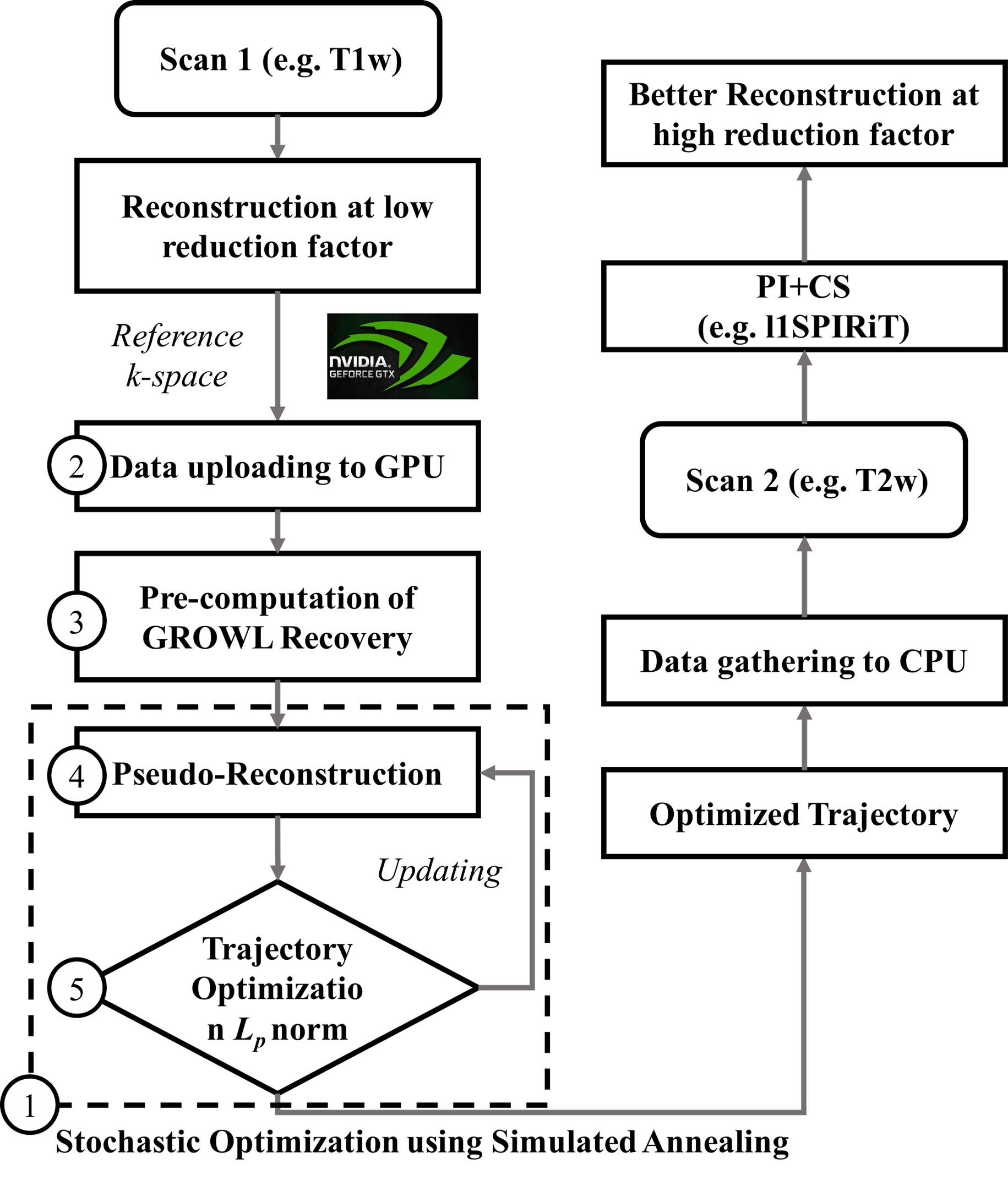}
\caption{Illustration of the flowchart of our proposed method. Sampling trajectory is optimized using Stimulated Annealing. This method is highly efficient due to pre-computation and GPU acceleration.} \label{Figure:fig1}
\end{figure}

\section{Theory}
\subsection{Under-sampling optimization for minimized reconstruction error}
Reconstruction of undersampled dataset is an inverse problem, which is to optimize image $x$ from measurement $y=UF(x)$ with Fourier encoding operator $F$ and undersampling operator $U$. The optimization of undersampling strategy is to in turn optimize $U$ to reach the minimized reconstruction error between reconstruction $\hat{x}(U)$ and the ground-truth image $x$ given certain sampling cost such as the number of independent phase-encoding samples.

\begin{equation}
U_{optimized} = \arg \min_{U}  { L(\hat{x}(y),x)}
\end{equation}



\subsection{Cost function for undersampling optimization}
Ideally, we want to optimize $U$ directly based on the reconstruction error, e.g. L2 metrics $||\hat(x)(UF(x))-x||_2^2$. Given the reconstruction itself is computational-intensive and there are large number of possible undersampling, even a greedy method with couples of reconstruction steps can be too slow to use in real applications. So an alternative surrogate cost function is required to approximate the reconstruction performance and guide the optimization. In this work, we propose to exploit the prior information, including sensitivity and anatomical priors, from multi-contrast sequences and to optimize undersampling from existing reference images for a new scan with similar coil sensitivity, anatomy but likely a different contrast. To combine both the prior knowledge of parallel imaging and compressed sensing, here we introduce a new surrogate function approximating PI+CS reconstruction errors.

We approximated reconstruction error using a modified coherence metric with Generalized Autocalibrating Partially Parallel Acquisitions (GRAPPA) operator.

\begin{equation}
f_p(u) = || \Phi (x - x(u)) || _p , (p >>1)
\end{equation}

This approximated cost function is close to the final PI+CS reconstruction error since it can be viewed as an error metrics of an approximated sequential PI+CS reconstruction. It first consider the missing k-space samples can be partially recovered using PI. Also similar to coherence metrics, it considers the artifact incoherence for CS.
Besides, this metrics can efficiently computed with an GRAPPA extrapolate operator, which is highly parallelizable and independent for each phase-encoding line. Therefore, the GRAPPA operator results can actually be simply computed once and stored to directly use in the optimization without further time cost. 

\section{Methods and material}
\subsection{Trajectory Optimization Framework}
Here in this work, we proposed a data-driven undersampling optimization method, which optimizes trajectories for subsequent multi-contrast series based on information from a previous (e.g. T1w, GRE) scan. Before the optimization procedure,  the previous scanned data (e.g. T1w, GRE) is sampled from the same or similar anatomy as reference k-space in order to get the optimized sampling trajectories. For each imaging protocol, the optimized trajectories could be extended to the following multi-contras series. The framework as well as five important features of the proposed method are shown in Figure \ref{Figure:fig1} for trajectory optimization.

1) An iterative random optimization scheme (heuristic Simulated Annealing) is used for choosing the best potential trajectory among all the candidates. Simulated Annealing \cite{kirkpatrick1983optimization} is a stochastic strategy for global optimization,which has been widely used in approximating the global optimum in a large search space. In our algorithm, a modified version of simulated annealing is implemented with heuristic updating. In each iteration, the candidates consist of two parts, the first being the best half candidates chosen from the last iteration, while the second being the candidates generated using genetic algorithm. Both two parts will be computed together to get the best candidates for the following iterations. Iteratively, we could use this method to generate more candidates and have the optimized trajectory after a certain number of iterations, which could be used for multi-contrast future scans.

2) A GRAPPA-operator\cite{griswold2002generalized} is pre-computed according to the reference data, and the k-space extrapolation using GRAPPA-operator is pre-calculated in each iteration.

3) A fast pseudo-reconstruction is implemented in each iteration to estimate the reconstruction performance, which would be much more efficient than the time consuming reconstruction procedure using PI+CS. $L_{p}$ Norm of errors in image domain is used as an objective function to balance PI and CS.

4) The performances of different trajectories are evaluated with the same reconstruction method using PI+CS method in The Berkeley Advanced Reconstruction Toolbox (BART) for comparison\cite{tamir2016generalized}. To asses the reconstruction performance, root-mean-square errors (RMSE) and the error maps between the fully-sampled and the reconstructed images are calculated for comparison.

5) The optimized trajectory calculated from previous scanned reference data (e.g. T1w, GRE) could be effectively used in future multi-contrast scans according to the sharable information between multi-dcontrast MRI datasets\cite{gong2015promise,li2012correlation}. 

To further improve the performance and efficiency of the method, both hardware acceleration (GPU )and software/algorithm level designs were used. Here we use a Nvidia GPU (GTX 1080Ti) for acceleration. The proposed GPU accelerated trajectory optimization algorithm as described in the previous section was implemented in the MATLAB 2016b programming environment with a 2.7 GHz Inter Core i7 CPU and 16 GB RAM.

\subsection{Datasets and experiments}
\subsubsection{Multi-Contrast brain imaging experiments}

To evaluate the performance and efficiency of the optimized trajectory, we conducted experiments to optimize trajectory using previously acquired T1w brain images and applied the optimized trajectory to Magnetic Resonance Compilation (MAGiC) \cite{MAGiC} images from which synthetic images with different contrast weightings can be acquired and generated on a 3T scanner (GE MR750 Waukesha, WI) using 12-channel head coil (HNS). 
For comparison, we compared the resulting errors using PI+CS reconstruction method (BART toolbox\cite{tamir2016generalized}) with different 1D-undersampling trajectory using various undersampling strategy at the same reduction factor ${\textmd{R}}$: 1) uniform undersampling. 2) variable density undersampling (ARC)\cite{beatty2007method}. 3) Point-Spread-Function (PSF) optimized pseudo-random trajectory. 4) the proposed method with random initialization. 5) the proposed method with ARC based initialization.  To evaluate the performance of the model, the Root-Mean Square Errors (RMSEs) between the images reconstructed from the optimized trajectory and the images reconstructed from the fully sampled k-space are calculated to investigate the performance of different trajectories. 
\subsubsection{Multi-Contrast Prostate Imaging Experiments}
Multi-Contrast prostate images were acquired with a 32-channel torsol coil on a 3T scanner (GE MR750 Waukesha, WI). 2D cartesian Gradient echo imaging was performed to obtain T1w images, with TE = 5.62ms, TR = 211.21ms, FOV = 24cm${\times}$24cm, in-plane resolution = 0.94mm${\times}$0.94mm and flip angle = 70${^\circ}$. Flow compensation was applied to avoid motion artifact from blood vessels. 
T2w images were acquired by a Fast Spin-Echo sequence with 23 echo trains. Image parameters are TE = ms, TR = 3500ms, FOV = 24cm${\times}$24cm and in-plane resolution = 0.94mm${\times}$0.94mm. Flow compensation was applied in the slice direction. 


\section{Results}
\subsubsection{Multi-Contrast Brain Imaging}
\begin{figure}[ht!]
\centering
\includegraphics[width=12cm]{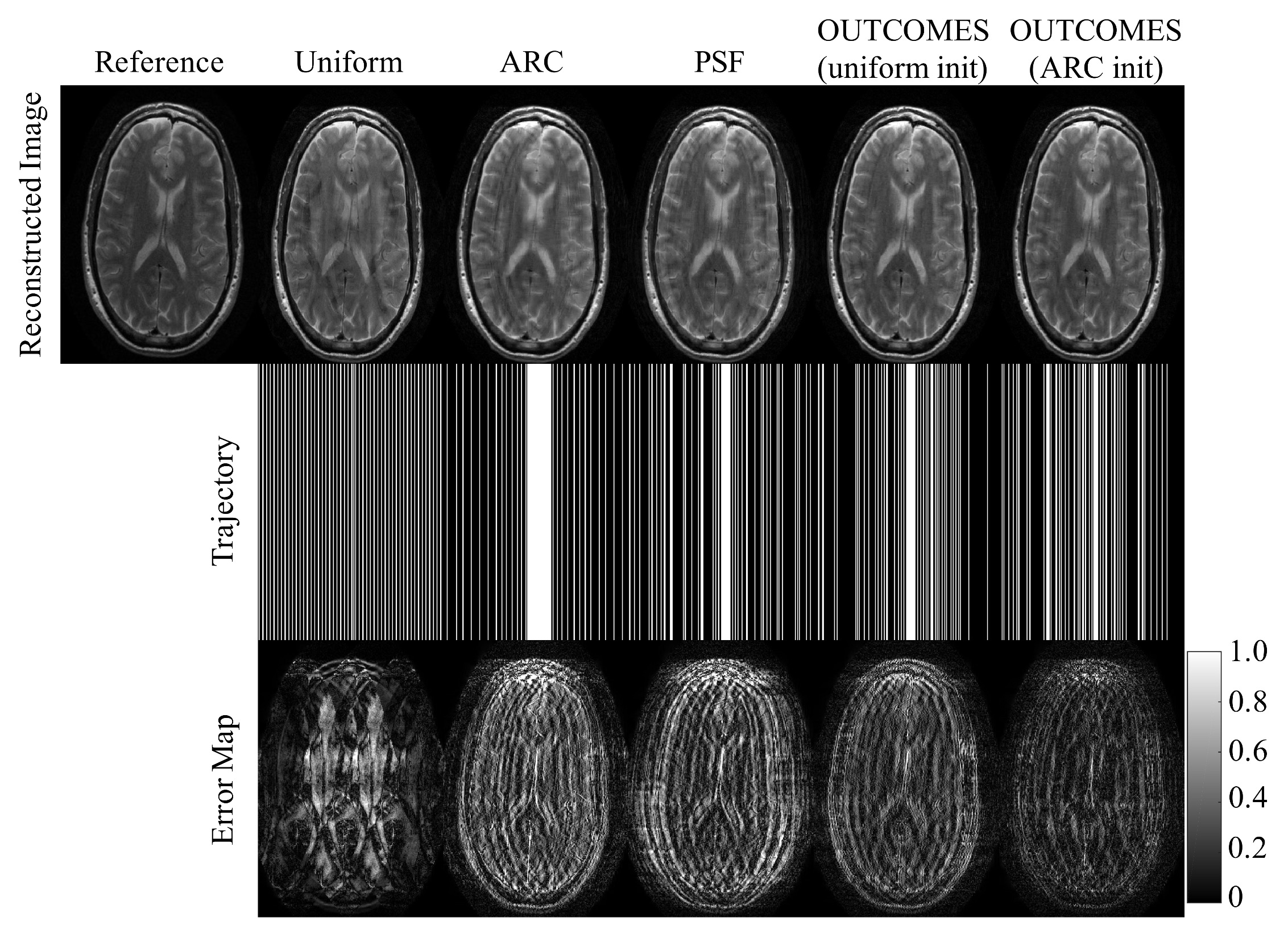}
\caption{Representative reconstructed images, undersampling trajectories and error maps from inversion 1 echo 1 of brain MAGiC acquisitions.} \label{fig:TrajEg}
\end{figure}

Figure \ref{fig:TrajEg} shows an example of the optimized undersampling trajectory from the proposed OUTCOMES method with both uniform undersampling initialization and ARC undersampling initialization. The reconstructed images and error maps from the proposed method are compared to the reference fully-sampled image and images from other undersampling patterns (ie. Uniform, ARC and PSF). It can be clearly seen that with a constant undersampling rate (R=4), OUTCOMES could largely reduce the artifacts and get a better 
reconstructed image quality.

\begin{figure}[ht!]
\centering
\includegraphics[width=10cm]{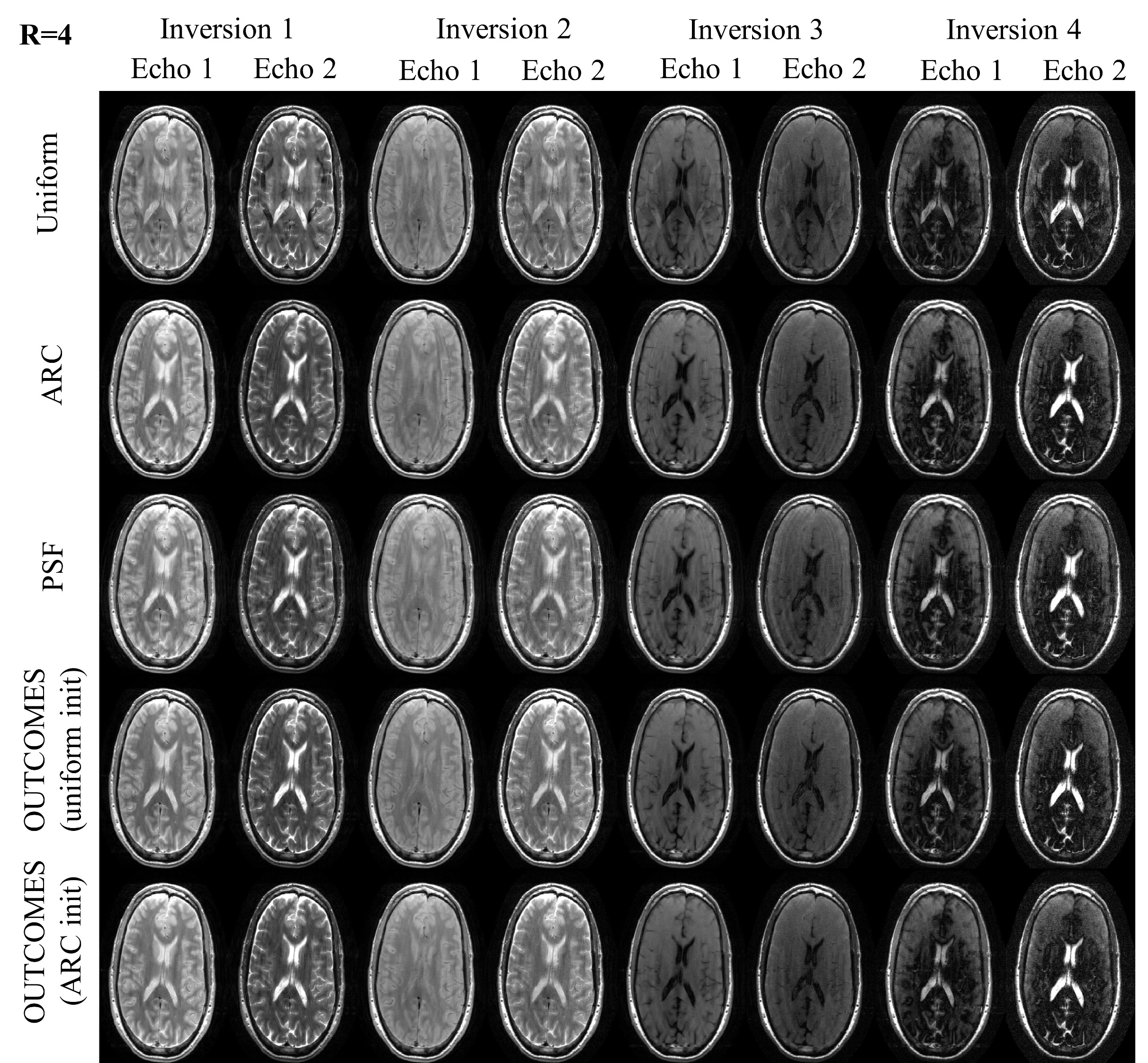}
\caption{Representative reconstructed images from different echos of brain MAGiC acquisitions using different reconstruction methods (Uniform, ARC, PSF, OUTCOMES (uniform init), and OUTCOMES (ARC init))} \label{fig:brainimgs}
\end{figure}

Figure \ref{fig:brainimgs} shows the comparison of the reconstructed eight contrast images of the MAGiC datasets using different trajectories (R=4): uniform undersampling, ARC, PSF optimized undersampling and the proposed OUTCOMES method with different initializations. For the proposed method, parameter set (20,50), which makes the algorithm to use 20 iterations with 50 candidates updated per iteration, was selected. Severe parallel imaging artifacts can be detected from Uniform, ARC and PSF, which are less obvious in images acquired by OUTCOMES. Table \ref{Tab:Hyper} shows the comparison of time costs for trajectory optimization and resulting reconstruction errors. To optimize a 1D random undersampling trajectory with the proposed method, it only takes 7.43 seconds on a GPU with improved matrix computation, while the traditional methods takes 54.91 seconds. The trade-off between candidate size and iteration number is shown in the Table \ref{Tab:Hyper}, where using 20~30 iteration and 30~50 candidate size would result in the best performance, which achieves the balance between time cost and improvements in resulting reconstruction using the optimized trajectory. Figure \ref{fig:S1} shows the RMSE results of different undersampling trajectories over increasing reduction factors for the images of eight brain contrasts. The RMSE of all methods is increasing with reduction factors, but the increasing rate for the proposed OUTCOMES method is slower than other methods in all the eight contrasts, which could demonstrate the efficiency and effectiveness of our methods when applied to multicontrast datasets.

\begin{table}[ht!]
    \centering
        \begin{center}
        \begin{tabular}{ c|c|c|c|c } 
        \hline
        Iterations & 10 & 20 & 33 & 50\\
        \hline
        Candidates & 100 & 50 & 33 & 20 \\ 
        Computation time (CPU)& 65.62s & 59.37s & 54.91s & 75.61s \\ 
        Computation time (GPU accelerated)& \textbf{10.71s} & \textbf{8.21s} & \textbf{7.43s} & \textbf{13.04s} \\ 
        RMSE of MAGiC contrast 1& 9.2\% & 6.1\% & 7.6\% & 6.4\% \\ 
       
        \hline
       
        \end{tabular}
        \end{center}
    \caption{Comparison of different parameter selesctions. For a fixed total number of 1000 candidate trajectories to evaluate, we tried multiple choices of parameter sets, including different iteration number and the number of candidate trajectories updated in each iteration: (10$\times$100), (20$\times$50), (33$\times$33), (50$\times$20).}
    \label{Tab:Hyper}
\end{table}

\subsubsection{Multi-Contrast Prostate Imaging}
\begin{figure}[ht!]
\centering
\includegraphics[width=12cm]{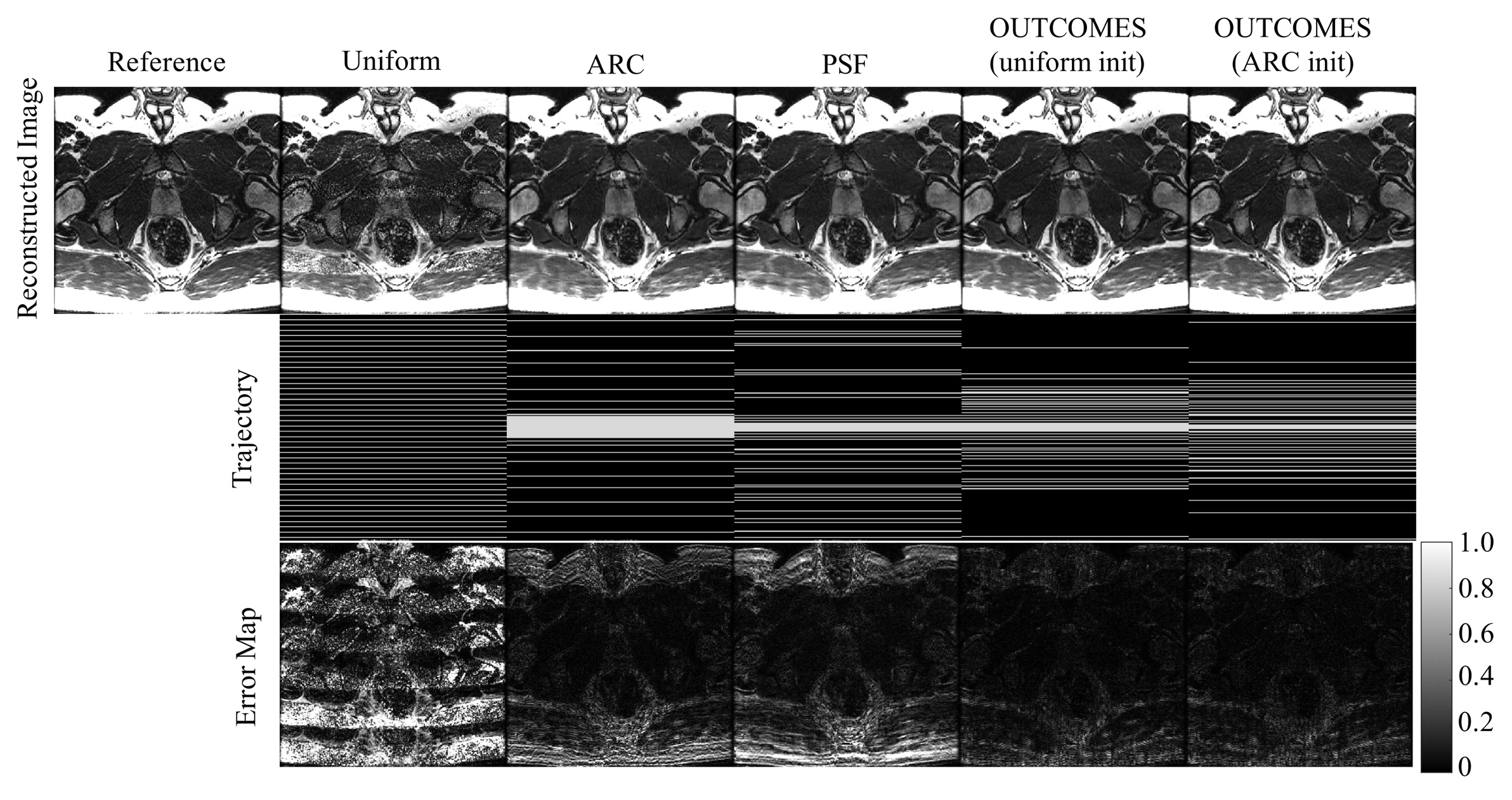}
\caption{Representative reconstructed images, undersampling trajectories and error maps from T2w prostate acquisitions.} \label{fig:prostate}
\end{figure}

Figure \ref{fig:prostate} shows the reconstructed images of T2w prostate dataset using different trajectories (R=6): uniform undersampling, ARC, PSF optimized undersampling and the proposed OUTCOMES method with both uniform initialization as well as ARC initialization. The optimized trajectories are obtained by implementing our method on T1w dataset from the same anatomy. At a higher reduction factor, Uniform undersampling would results in sever aliasing artifacts which are not obvious in images reconstructed by OUTCOMES.

\section{Conclusions}
In this work, we proposed an efficient trajectory optimization framework: OUTCOMES, which uses GPU accelerated efficient pesudo reconstruction instead of complicated reconstruction, combined with redesigns and acceleration, can be over 10x faster than original un-accelerated version and over 100x faster than more sophisticated trajectory optimization methods. Besides, the optimized trajectory obtained from a single contrast could also be well applied to different contrast, further saving the computation time and accelerate the scanning process. 
%
%
%
%
\clearpage
\newpage
\bibliographystyle{splncs04}
\bibliography{ref}
\clearpage
\newpage
\section{Supplementary Material}
\renewcommand{\thefigure}{S1}
\begin{figure}
\centering
\includegraphics[width=11cm]{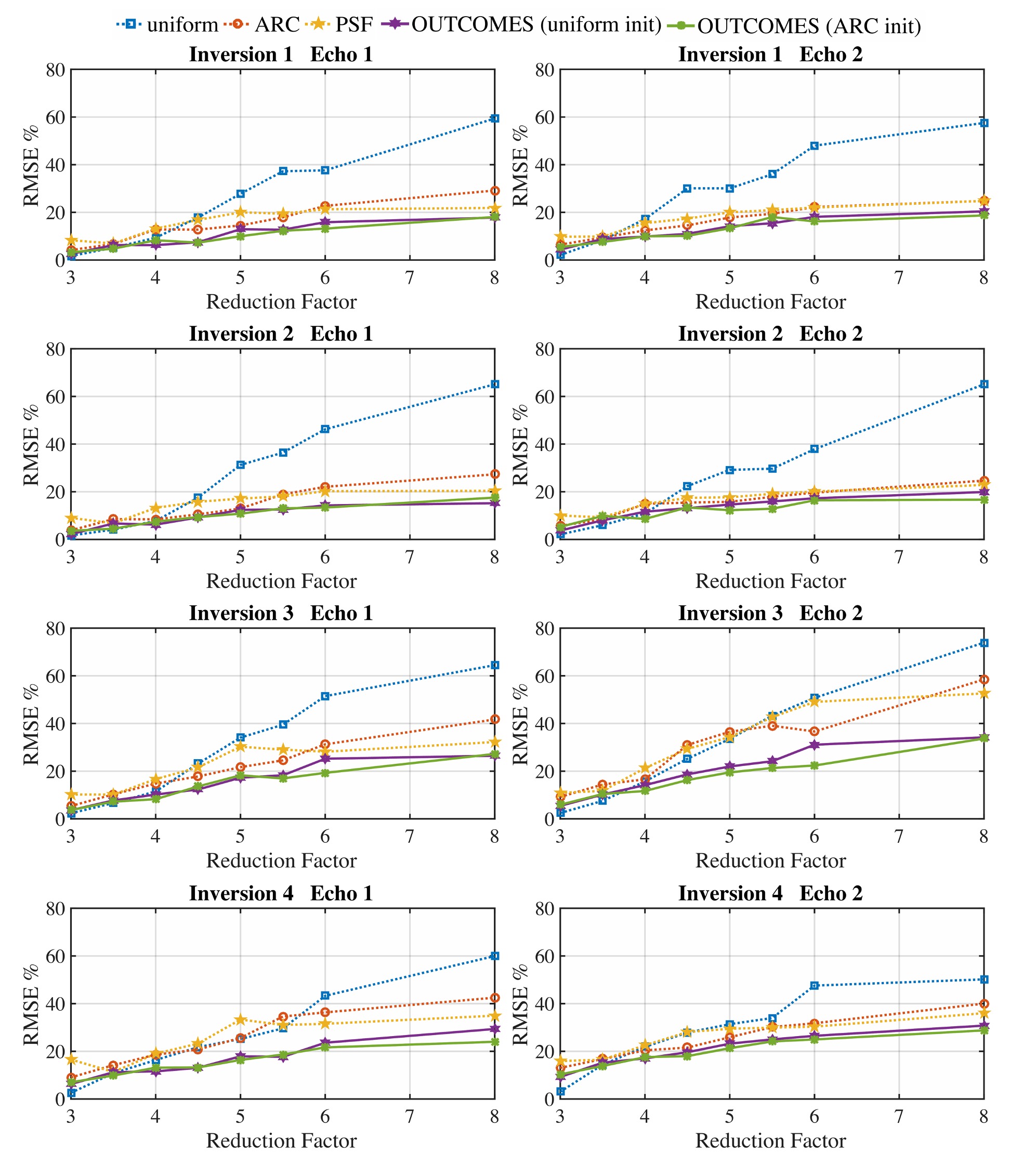}
\caption{Quantitative NRMSE comparison at different undersampling reduction factor R for MAGiC acquisition (8 contrasts). Different reconstruction methods (Uniform, ARC, PSF, OUTCOMES (uniform init) and OUTCOMES (ARC init)) are included for comparisons.
} 
\label{fig:S1}
\end{figure}


\end{document}